\documentclass[prd,onecolumn,notitlepage,eqsecnum,nofootinbib,floatfix]{revtex4-1}
\usepackage{amssymb}
\usepackage{amsmath}
\usepackage{amsfonts} 
\usepackage{bm}
\usepackage{graphicx}
\usepackage{comment} 
\DeclareSymbolFont{EulerScript}{U}{eus}{m}{n}
\SetSymbolFont{EulerScript}{bold}{U}{eus}{b}{n}
\DeclareSymbolFontAlphabet\scrpt{EulerScript}
\newcommand{\unitr}{\bm{\hat{r}}} 
\newcommand{\unittheta}{\bm{\hat{\theta}}} 
\newcommand{\unitphi}{\bm{\hat{\phi}}} 
\newcommand{\unitx}{\bm{\hat{x}}} 
\newcommand{\unity}{\bm{\hat{y}}} 
\newcommand{\unitz}{\bm{\hat{z}}} 
\newcommand{\gothj}{\mathfrak{j}}
\newcommand{\gothm}{\mathfrak{m}}
\newcommand{\m}{{\sf m}} 
\allowdisplaybreaks
\begin{document}
\title{Self-torque and angular momentum balance for a spinning charged
sphere} 
\author{B\'eatrice Bonga} 
\affiliation{Perimeter Institute for Theoretical Physics, Waterloo,
  Ontario N2L 2Y5, Canada} 
\author{Eric Poisson}  
\affiliation{Department of Physics, University of Guelph, Guelph,
  Ontario, N1G 2W1, Canada} 
\author{Huan Yang} 
\affiliation{Department of Physics, University of Guelph, Guelph,
  Ontario, N1G 2W1, Canada}
\affiliation{Perimeter Institute for Theoretical Physics, Waterloo,
  Ontario N2L 2Y5, Canada} 
\date{August 27, 2018} 
\begin{abstract} 
Angular momentum balance is examined in the context of the
electrodynamics of a spinning charged sphere, which is
allowed to possess any variable angular velocity. We calculate 
the electric and magnetic fields of the (hollow) sphere, and express
them as expansions in powers of $\tau/t_c \ll 1$, the ratio of the
light-travel time $\tau$ across the sphere and the characteristic time
scale $t_c$ of variation of the angular velocity. From the fields we
compute the self-torque exerted by the fields on the sphere, and argue
that only a piece of this self-torque can be associated with radiation
reaction. Then we obtain the rate at which angular momentum is
radiated away by the shell, and the total angular momentum contained
in the electromagnetic field. With these results we demonstrate
explicitly that the field angular momentum is lost in part to
radiation and in part to the self-torque; angular momentum balance is
thereby established. Finally, we examine the angular motion of the
sphere under the combined action of the self-torque and an additional
torque supplied by an external agent.     
\end{abstract} 
\maketitle

\section{Introduction and summary} 
\label{sec:intro} 

The aim of this paper is to investigate the electrodynamics of a
spinning charged sphere. Our wish is to bring together two
themes of the standard textbook literature, one rather neglected, 
the other shrouded in mystery.      

The neglected theme is electromagnetic angular momentum, which tends
to be mentioned in passing in most texts, but rarely explored beyond
a very small number of examples. For example, Griffiths
\cite{griffiths:13} introduces the field angular momentum in
Sec.~8.2.4 and computes it explicitly for a solenoid encased between
two cylindrical shells.\footnote{This variant of Feynman's disk
  paradox was first proposed by Romer \cite{romer:66} and Boos
  \cite{boos:84}; the original paradox is discussed in Sec.~17.4 of
  Volume II of the Feynman lectures \cite{feynman-leighton-sands:11},
  and in Ref.~\cite{lombardi:83}.} Apart from another appearance in
Problem 8.17, this is the sole mention of electromagnetic angular
momentum in this popular text. As another example, Jackson
\cite{jackson:99} mentions the field angular momentum on a few
occasions, but the topic is not deemed worthy of a section of its own.    

The theme shrouded in mystery is self-force and radiation
reaction. Despite valiant attempts to elucidate this topic 
in the standard texts (see, for example, Sec.~11.2.2 in Griffiths, or
Chapter 16 of Jackson), these aspects of electromagnetic theory remain 
conceptually challenging to most students (and instructors), thanks to
the lurking infinities, mass renormalization, runaway solutions,
pre-acceleration, and the like. This topic has been the subject of a
vast literature, and {\sl American Journal of Physics} has been an
active participant. A simple pedagogical introduction to the subject
can be found in Boyer's paper \cite{boyer:72}, and a useful historical
survey was provided by Rohrlich \cite{rohrlich:97}. In
Ref.~\cite{griffiths-proctor-schroeter:10} Griffiths, Proctor, and
Schroeter compare the predictions of the standard equation of motion
for a charged particle, the Abraham-Lorentz equation, to those of the
less pathological Landau-Lifshitz equation (see Sec.~75 of
Ref.~\cite{landau-lifshitz:b2}). The issue of mass renormalization is
discussed by Griffiths and Owen \cite{griffiths-owen:83}. A clever
derivation of the self-force was devised by Griffiths and Szeto 
\cite{griffiths-szeto:78} on the basis of a dumbbell model for an
electric charge; their construction was revisited and generalized by
Ori and Rosenthal \cite{ori-rosenthal:03, ori-rosenthal:04}.  

In our opinion, the conceptual difficulties associated with self-force
and radiation reaction stem from a misguided insistence that the
Maxwell-Lorentz theory should apply to a point particle. While the
electromagnetic field of a point particle is perfectly well defined,
its action back on the particle is not, and much contortion is
required to obtain self-consistent equations of motion for the
particle. Our point of view is that in the classical framework of the
Maxwell-Lorentz theory, equations of motion should only be formulated
for extended blobs of charge, and a pointlike limit should be
recovered only when the electromagnetic field varies on a length scale
that is long compared with the size of the blob. In this context the
equations of motion are necessarily approximate, and self-consistency
can be achieved only within the limits of the approximation. For a
rigorous implementation of this idea, see
Ref.~\cite{gralla-harte-wald:09}.   

Our interest in this paper is with the electrodynamics of an
infinitely thin, spherical shell that is uniformly charged and 
spinning on a central axis with a variable
angular velocity. The spinning motion of the sphere implies that
angular momentum shall be one of its most important attributes,
and thus is the first theme introduced.\footnote{The electromagnetic
angular momentum of a spinning shell with constant angular velocity
was previously calculated by de Castro \cite{decastro:91}.} The fact 
that the angular velocity is variable implies that the shell emits
electromagnetic radiation, which takes angular momentum away from the 
system; this loss must be balanced by a self-torque acting on the
shell, and we have our second theme. Such issues were explored
previously in a paper by Stump and Pollack \cite{stump-pollack:97}, who
calculated the self-torque acting on a point magnetic dipole. We feel
that the spinning shell, with its finite extension, provides a much more
satisfactory starting point for this discussion.     

The system's angular-momentum vector $\bm{J}$ is subjected to two
conservation equations, which are established in
Sec.~\ref{sec:conservation}. The first concerns $\bm{J}^{\rm em}$, the
electromagnetic piece of the angular momentum, and it reads  
\begin{equation} 
\frac{d}{dt} \bm{J}^{\rm em} =-\bm{\scrpt N}^{\rm em} 
- \bm{\scrpt T}^{\rm em}, 
\label{cons1} 
\end{equation} 
where $\bm{\scrpt N}^{\rm em}$ is the rate at which angular momentum
is radiated away, and $\bm{\scrpt T}^{\rm em}$ is the electromagnetic
torque --- the self-torque --- acting on the shell. The equation
states that the field angular momentum is communicated partly to the
radiation and partly to the shell. The second conservation statement
concerns $\bm{J}^{\rm mech}$, the shell's mechanical angular momentum,
and it reads 
\begin{equation} 
\frac{d}{dt} \bm{J}^{\rm mech} = \bm{\scrpt T}^{\rm em} 
+ \bm{\scrpt T}^{\rm ext},  
\label{cons2} 
\end{equation} 
where $\bm{\scrpt T}^{\rm em}$ is again the self-torque, and 
$\bm{\scrpt T}^{\rm ext}$ is the torque supplied by an external agent
acting on the shell. The equation states that the shell's angular
momentum changes because of the combined action of both torques. The
sum of $\bm{J}^{\rm em}$ and $\bm{J}^{\rm mech}$ is the total angular
momentum $\bm{J}$, and according to Eqs.~(\ref{cons1}) and
(\ref{cons2}), it satisfies the conservation statement 
$d\bm{J}/dt = -\bm{\scrpt N}^{\rm em} - \bm{\scrpt T}^{\rm ext}$.  
The total angular momentum is not conserved, because of radiative
losses and the torque exerted by the external agent.   

The conservation statement of Eq.~(\ref{cons1}) can be verified by
calculating the field angular momentum, the radiated losses, and the
self-torque. This requires expressions for the shell's electric and
magnetic fields at any time $t$ and any position $\bm{x}$. Provided
that we introduce a mild and sensible assumption, the fields of a 
spinning charged sphere can be calculated for any variable angular
velocity. The assumption is that the angular velocity is taken to vary
on a characteristic time scale $t_c$ that is very long compared with
$\tau = R/c$, the light-travel time across (half) the sphere; $R$ is
the sphere's radius and $c$ is the speed of light. The fields are
calculated in Sec.~\ref{sec:fields} and presented as expansions
in powers of $\tau/t_c \ll 1$. In principle, these can be carried out 
to an arbitrarily high order; we truncate them after a representative 
number of terms. 

With the fields in hand, we calculate the self-torque in
Sec.~\ref{sec:self-torque}, and express it also as an expansion in 
powers of $\tau/t_c$. We find that only pieces of
the self-torque can be associated with radiation reaction; the
remaining pieces (which include the dominant contribution to the
self-torque) have nothing to do with radiation reaction. As we explain
in Sec.~\ref{sec:self-torque}, the self-torque can be decomposed into
terms that come with even powers of $\tau/t_c$ (and therefore even
powers of $c$), and terms that come with odd powers of $\tau/t_c$ (and
therefore odd powers of $c$). The odd terms are sensitive to the
choice of boundary condition at infinity; they change sign when the
retarded solution describing an outgoing wave is switched to the
advanced solution describing an incoming wave. These odd terms,
therefore, can be associated with radiation reaction, with the
justification that a switch of boundary condition will produce a change
in the self-torque. The even terms, on the other hand, are insensitive
to the choice of boundary condition, and stay the same after a switch;
these terms have nothing to do with radiation reaction. The dominant
contribution to the self-torque is even in $\tau/t_c$.      

The radiated angular momentum is calculated in
Sec.~\ref{sec:radiated}, and presented as an expansion in powers of
$\tau/t_c$. As can be expected from the preceding discussion, each
term is odd in $c$, and therefore sensitive to the choice of boundary
condition at infinity. We note that the radiated angular momentum
(like the self-torque) is proportional to $q$, the total charge of the
spinning sphere. This provides a vivid illustration of the fact that
the flux of angular momentum depends not only on the radiative, $1/r$
part of the electromagnetic field, but also on its Coulombic, $1/r^2$
part; this point was made forcibly in a recent paper by Ashtekar and
Bonga \cite{ashtekar-bonga:17}. The field's angular momentum is
calculated in Sec.~\ref{sec:angmoment}, and Eq.~(\ref{cons1}) is
verified explicitly.  

In Sec.~\ref{sec:motion} we turn to the shell's angular motion,
governed by Eq.~(\ref{cons2}). We imagine that an external agent
applies a torque on the shell, in addition to the self-torque supplied
by the electromagnetic field. The resulting equation of motion is
again presented as an expansion in powers of $\tau/t_c$, with the
external torque giving the dominant contribution, and the self-torque
splitting up into radiation-reaction terms (odd in $\tau/t_c$) and
terms (even in $\tau/t_c$)  that have nothing to do with radiation
reaction. A feature of the equation of motion is that it is written in
terms of an effective mass $M_{\rm eff}$, the sum of the shell's mass
and the additional inertia contributed by the electrostatic field;
this is a form of mass renormalization, a ubiquitous procedure of
self-force calculations. Another feature of the equation
of motion is that it requires a reduction of order (see Sec.~75 of
Ref.~\cite{landau-lifshitz:b2} or Ref.~\cite{cook:84}) to avoid the
emergence of runaway solutions. In our case this procedure is natural
and justified on the grounds of the underlying assumption that
$\tau/t_c \ll 1$. We examine several examples of the shell's angular
motion, corresponding to different external torques.  

The electrodynamics of a spinning charged sphere unlocks a number of
important lessons. First, it presents an excellent opportunity to 
reflect on the role of angular momentum in electromagnetism, a largely
neglected subject. Second, it provides a vivid example of the fact
that the radiated angular momentum may not solely depend on the $1/r$, 
radiative part of the electromagnetic field, but can also depend on the
$1/r^2$, Coulombic part. Third, it provides a conceptually clean setup
to understand how the field can act back on its source; the self-torque
is well defined and can be computed explicitly, and this should help
lift some of the aura of mystery that tends to accompany this topic. 
Fourth, it reveals that the notions of self-torque and radiation
reaction must be clearly distinguished. Indeed, the self-torque
contains two types of terms: time-antisymmetric terms that reflect the
choice of outgoing-wave boundary condition and can therefore be
associated with radiation reaction, and time-symmetric terms that are
insensitive to the choice of boundary condition and have therefore
nothing to do with radiation reaction. And fifth, it provides an
instance in which the equation of motion for the source of an
electromagnetic field is well defined and can be solved
explicitly. These lessons can all be learned at a reasonable cost: All
calculations are straightforward and accessible to a starting graduate
student, and they might also be tractable to a well-educated senior
undergraduate (instructors should be able to cope as well). We believe
that these lessons are well worth learning, and that this paper could
form the basis of an independent study module in a graduate course in
electromagnetism.         
 
This paper is dedicated to the memory of Steve Detweiler, who
understood better than most the power of simple models to foster
understanding of complicated things. 

\section{Conservation statements}
\label{sec:conservation} 

In this section we establish a number of conservation statements
relating to the momentum and angular momentum of an electromagnetic
field. These are formulated in some generality, with the assumption
that the field is sourced by a bounded charge distribution supported
by a fluid; for simplicity we take the fluid to be electrically and
magnetically unpolarized. The conservation laws will then apply, as a
special case, to the field sourced by a spinning charged sphere. 

Throughout the paper we denote three-dimensional vectors with a
bold-italic symbol such as $\bm{E}$, and denote individual components
with $E_a$ or $(\bm{E})_a$. We shall move freely between the
notations, making the most convenient choice given the context. We
adopt the Einstein summation convention, according to which repeated
vector (or tensor) indices are summed over. As examples of our
notation, we write $|\bm{E}|^2 = \bm{E} \cdot \bm{E} = E_a E_a$ and 
$(\bm{E} \times \bm{B})_a = \epsilon_{abc} E_b B_c$, where 
$\epsilon_{abc}$ is the completely antisymmetric permutation
symbol. We let $\partial_t := \partial/\partial t$ stand 
for partial differentiation with respect to time $t$, and $\partial_a
:= \partial/\partial x^a$ for partial differentiation with respect to
the spatial coordinate $x^a$.   

We begin with a conservation statement for the linear momentum
contained in the electromagnetic field. Let 
\begin{equation} 
g_a^{\rm em} := \epsilon_0 (\bm{E} \times \bm{B})_a 
\end{equation} 
be the field's momentum density, and 
\begin{equation} 
T^{\rm em}_{ab} := -\epsilon_0 \Bigl( E_a E_b  
- \frac{1}{2} \delta_{ab} |\bm{E}|^2 \Bigr)
- \frac{1}{\mu_0} \Bigl( B_a B_b  
- \frac{1}{2} \delta_{ab} |\bm{B}|^2 \Bigr) 
\label{maxwell_stress} 
\end{equation} 
be the field's stress tensor, defined such that $T^{\rm em}_{ab}\, dS_b$
is the momentum crossing out of an element of surface $dS_b$ per unit 
time.\footnote{The stress tensor is sometimes defined with plus
  signs, with the alternative convention that $T^{\rm em}_{ab}\, dS_b$
  is the momentum crossing {\it into} the element of surface per unit time.}
Finally, let  
\begin{equation} 
f^{\rm em}_a := \rho E_a + (\bm{j} \times \bm{B})_a 
\label{force_em} 
\end{equation} 
be the Lorentz force density on a charge distribution described by
the charge density $\rho$ and current density $\bm{j}$. Then the
conservation statement  
\begin{equation} 
\partial_t g_a^{\rm em} + \partial_b T^{\rm em}_{ab} = -f^{\rm em}_a 
\label{momentum_thm} 
\end{equation} 
follows as a consequence of Maxwell's equations. This is the momentum
theorem for the electromagnetic field, which is proved in the standard
texts (see, for example, Sec.~8.2.2 of Griffiths \cite{griffiths:13} or
Sec.~6.7 of Jackson \cite{jackson:99}). 

The momentum theorem gives rise to an angular-momentum theorem, which
we formulate in integral form. Let $V$ be a region of
three-dimensional space bounded by a closed two-surface $S$. We assume 
that $V$ and $S$ are fixed in space. Let 
\begin{equation} 
J^{\rm em}_a := \int_V ( \bm{x} \times \bm{g}^{\rm em} )_a\, d^3x 
\label{em_angmom} 
\end{equation} 
be the total angular momentum of the electromagnetic field in $V$;
$\bm{x}$ denotes the position vector. Then a consequence of
Eq.~(\ref{momentum_thm}) is the conservation statement 
\begin{equation} 
\frac{d}{dt} J^{\rm em}_a = -{\scrpt N}^{\rm em}_a 
- {\scrpt T}_a^{\rm em}, 
\label{conservation_em} 
\end{equation} 
where 
\begin{equation} 
{\scrpt N}^{\rm em}_a := \epsilon_{abc} \oint_S x_b T^{\rm em}_{cd}\, dS_d 
\label{flux} 
\end{equation} 
is the field angular momentum crossing out of $S$ per unit time, and 
\begin{equation} 
{\scrpt T}_a^{\rm em} := \epsilon_{abc} \int_V x_b f^{\rm em}_c\, d^3x 
\label{torque_em} 
\end{equation} 
is the torque exerted by the fields on the charge distribution in
$V$. Equation (\ref{conservation_em}) states that the rate at which
the field angular momentum leaves $V$ is equal to the rate at which it
crosses out of $S$ added to the torque exerted on all charges within
$V$. The equation is established by taking the time derivative
inside the integral in Eq.~(\ref{em_angmom}), substituting
Eq.~(\ref{momentum_thm}), and integrating by parts. 

The fluid that supports the charge distribution contributes
mechanical momentum and angular momentum. Let  
$\rho_{\rm m}$ be the fluid's mass density, and let $v_a$ be its
velocity field. Mass conservation is expressed by the continuity
equation $\partial_t \rho_{\rm m} + \partial_a (\rho_{\rm m} v_a) 
= 0$, and in the absence of charges and fields, the fluid is subjected
to the momentum theorem  
\begin{equation} 
\partial_t g^{\rm mech}_a + \partial_b T^{\rm mech}_{ab} 
= f^{\rm ext}_a, 
\end{equation} 
where $g^{\rm mech}_a = \rho_{\rm m} v_a$ is the mechanical momentum
density, $T^{\rm mech}_{ab}$ the mechanical stress tensor, and
$f^{\rm ext}_a$ the force density supplied by external agents. In
the case of a perfect fluid, for example, the mechanical stress tensor
is given by $T^{\rm mech}_{ab} = \rho_{\rm m} v_a v_b 
+ p\, \delta_{ab}$, where $p$ is the pressure, and the momentum theorem
becomes $\rho_{\rm m} dv_a/dt + \partial_a p = f^{\rm ext}_a$. This
is recognized as Euler's equation. 

If we now allow the fluid to be charged, and allow for the presence of
an electromagnetic field, the momentum theorem becomes 
\begin{equation} 
\partial_t g_a + \partial_b T_{ab} = f^{\rm ext}_a, 
\end{equation} 
where $g_a := g^{\rm mech}_a + g^{\rm em}_a$ is the total momentum
density, $T_{ab} := T^{\rm mech}_{ab} + T^{\rm em}_{ab}$ the total
stress tensor, and $f^{\rm ext}_a$ the force density provided by
external agents (excluding the Lorentz force density). In the case of
a perfect fluid, the momentum theorem becomes $\rho_{\rm m} dv_a/dt
+ \partial_a p - f^{\rm em}_a = f^{\rm ext}_a$, Euler's equation for a
charged fluid. 

The momentum theorem for a charged fluid gives rise to a conservation
statement for the total angular momentum contained in the region $V$,  
\begin{equation} 
J_a := \int_V (\bm{x} \times \bm{g})_a\, d^3x 
= J_a^{\rm mech} + J_a^{\rm em}.
\end{equation} 
We assume that the fluid occupies a bounded region within $V$, and
that this region does not extend all the way to $S$. In this case
$T_{ab}$ evaluated on $S$ consists of a field contribution only, and
the mechanical contribution vanishes. The conservation statement is 
\begin{equation} 
\frac{d}{dt} J_a = -{\scrpt N}^{\rm em}_a - {\scrpt T}_a^{\rm ext}, 
\label{conservation_total} 
\end{equation} 
where ${\scrpt N}^{\rm em}_a$ is the field angular momentum crossing
out of $S$ per unit time, as given by Eq.~(\ref{flux}), and 
\begin{equation} 
{\scrpt T}^{\rm ext}_a := \epsilon_{abc} \int_V x_b f^{\rm ext}_c\, d^3x 
\end{equation} 
is the torque supplied by the external agents. 
 
Equations (\ref{conservation_em}) and (\ref{conservation_total}) imply 
\begin{equation} 
\frac{d}{dt} J^{\rm mech}_a = {\scrpt T}_a^{\rm em} 
+ {\scrpt T}_a^{\rm ext}, 
\label{conservation_mech} 
\end{equation} 
the statement that the mechanical angular momentum changes because of
the combined action of the electromagnetic and external torques. 

\section{Fields of a spinning charged sphere} 
\label{sec:fields}  
 
We consider an infinitely thin, spherical shell of radius $R$ spinning
on a central axis with a time-changing angular velocity
$\Omega(t)$. The shell possesses a tangential pressure that supports
it against gravity, and we assume that its material is not polarized
(either electrically or magnetically) by the electromagnetic field.  
We align the $z$-direction with the rotation axis, and employ the
usual spherical coordinates $(r,\theta,\phi)$ in addition to the
Cartesian coordinates $(x,y,z)$. Each coordinate system comes with a
basis of unit vectors. For the Cartesian coordinates we have $(\unitx,
\unity, \unitz)$, and for the spherical coordinates we have $(\unitr,
\unittheta, \unitphi)$. 

The shell carries a charge $q$, and its charge
density is given by  
\begin{equation} 
\rho = \frac{q}{4\pi R^2} \delta(r-R). 
\end{equation} 
The velocity of an element of charge on the shell is $\bm{v} = R
\Omega(t) \sin\theta\, \unitphi$. The current density is 
$\bm{j} = \rho \bm{v}$, and the shell's magnetic moment is 
\begin{equation} 
\bm{m} := \frac{1}{2} \int \bm{x} \times \bm{j}\, d^3 x 
= \frac{1}{3} q R^2 \Omega(t)\, \unitz.  
\end{equation} 
In the following we shall express $\Omega(t)$ in terms of $m(t)$, with  
$m := |\bm{m}| = \frac{1}{3} q R^2 \Omega$. 

In the usual Lorenz gauge, the potentials created by the spinning
shell satisfy wave equations, with solutions 
\begin{subequations} 
\begin{align} 
\Phi(t,\bm{x}) &= \frac{1}{4\pi\epsilon_0} 
\int \rho(t',\bm{x'}) G(t,\bm{x};t',\bm{x'})\, dt' d^3x', \\ 
A_a(t,\bm{x}) &= \frac{\mu_0}{4\pi} 
\int j_a(t',\bm{x'}) G(t,\bm{x};t',\bm{x'})\, dt' d^3x', 
\end{align} 
\end{subequations} 
where $G(t,\bm{x};t',\bm{x'})$ is the retarded Green's function for
the wave equation. Because $\rho$ is static and spherically
symmetric, the scalar potential is given simply by 
\begin{subequations} 
\begin{align} 
\Phi_{\rm in} &= \frac{1}{4\pi\epsilon_0} \frac{q}{R}, \\ 
\Phi_{\rm out} &= \frac{1}{4\pi\epsilon_0} \frac{q}{r},
\end{align} 
\end{subequations} 
with $\Phi_{\rm in}$ applying when $r < R$, and $\Phi_{\rm out}$
when $r > R$. 

Because $j_a$ is time dependent (via $m$), the computation of the
vector potential is more involved. It is helpful to work with 
$A_y - i A_x$, noting that $j_z = 0$ and $j_y - i j_x 
\propto \sin\theta e^{i\phi}$, a spherical harmonic with labels 
$\ell = 1$ and $\m = 1$. We take advantage of a decomposition of the
Green's function in spherical harmonics, provided by 
\begin{equation} 
G(t,\bm{x};t',\bm{x'}) = \sum_{\ell \m} g_\ell(t,r;t',r')
Y^*_{\ell \m}(\theta',\phi') Y_{\ell \m}(\theta,\phi), 
\label{Gdecomp} 
\end{equation} 
where $(r,\theta,\phi)$ are the spherical coordinates attached to
$\bm{x}$, while $(r',\theta',\phi')$ are those attached to $\bm{x'}$,
and where 
\begin{equation} 
g_\ell(t,r;t',r') = \frac{2\pi c}{rr'} \Theta(\Delta - |r-r'|) 
\Theta(r+r'-\Delta) P_\ell(\xi), 
\label{g_def} 
\end{equation} 
with $\Theta(u)$ denoting the step function, $P_\ell(\xi)$ the Legendre
polynomials, $\Delta := c(t-t')$, and 
\begin{equation} 
\xi := \frac{r^2 + r^{\prime 2} - \Delta^2}{2rr'}. 
\end{equation} 
Because this decomposition does not seem to be widely known, but can
be immensely useful, we provide a derivation in the Appendix.  

We insert Eq.~(\ref{Gdecomp}) within the integral for $A_y 
- iA_x$, write $d^3 x' = r^{\prime 2} dr'\, d\Omega'$ where  
$d\Omega' := \sin\theta' d\theta' d\phi'$, perform the angular
integration by exploiting the orthonormality of spherical harmonics,
and evaluate the radial integral with the help of the delta function
contained in $\bm{j}$. We arrive at  
\begin{equation} 
\bm{A} = \frac{\mu_0}{4\pi} \Gamma(t,r) \sin\theta\, \unitphi, 
\label{A_result} 
\end{equation} 
where 
\begin{equation} 
\Gamma(t,r) := \frac{3}{4\pi R} \int m(t') g_1(t,r;t',R)\, dt'. 
\end{equation} 
Inserting Eq.~(\ref{g_def}) with $\ell=1$ and accounting for the step
functions, this function becomes 
\begin{equation} 
\Gamma(t,r) = \frac{3c}{4R^3 r^2} \int_{t-(r+R)/c}^{t-|r-R|/c} 
m(t') \bigl[ r^2 + R^2 - c^2 (t-t')^2 \bigr]\, dt'. 
\label{Gamma_def} 
\end{equation} 
In principle, the integral can be evaluated only once $m(t')$ is
specified. As we shall see, however, we can avoid making an explicit 
identification, and obtain $\Gamma(t,r)$ for any $m(t')$, at the small
price of a sensible assumption. We shall assume that $t_c$, the
characteristic time scale of variation of $m(t')$, is very long
compared with $\tau := R/c$, the light-travel time across (half) the
sphere. The assumption implies, for example, that 
$\tau \dot{m} \ll m$, that $\tau^2 \ddot{m} \ll \tau \dot{m}$, and so
on, with overdots indicating differentiation with respect to time. 

When $r < R$, the bounds of integration in Eq.~(\ref{Gamma_def})
become $t-R/c-r/c$ and $t-R/c+r/c$, respectively, and the integration
takes place over a short time interval $2r/c < 2\tau$ around
$t=R/c$. Over this interval $m(t')$ can be expressed as a Taylor
expansion about $t' = t-R/c$, and the integral can be evaluated. We
obtain  
\begin{equation} 
\Gamma_{\rm in}(t,r) = \frac{r}{R^3} \biggl[ 
\bigl( m + \tau \dot{m} \bigr) 
+ \frac{1}{10} \frac{r^2}{c^2} \bigl( m^{(2)} + \tau m^{(3)} \bigr) 
+ \frac{1}{280} \frac{r^4}{c^4} \bigl( m^{(4)} + \tau m^{(5)} \bigr) 
+ \frac{1}{15120} \frac{r^6}{c^6} \bigl( m^{(6)} + \tau m^{(7)} \bigr) 
+ \cdots \biggr]_{t-R/c}, 
\label{Gamma_in} 
\end{equation} 
where a number within brackets in a superscript gives the number
of differentiations, and the subscript after the end square bracket
indicates that $m$ and its derivatives are evaluated at
the time $t-R/c$. We remark that $m$ and its derivatives could be
further Taylor expanded about the time $t$, but this form for 
$\Gamma_{\rm in}$ happens to be convenient to work with. 

When $r > R$, the bounds of integration in Eq.~(\ref{Gamma_def})
become $t-r/c-R/c$ and $t-r/c+R/c$, respectively, and the integration
again takes place over a short time interval $2R/c = 2\tau$. Over this
interval $m(t')$ can again be expressed as a Taylor expansion, this
time about $t' = t-r/c$, and the integral can again be evaluated. We
obtain  
\begin{equation} 
\Gamma_{\rm out}(t,r) = \frac{1}{r^2} \biggl[ 
\Bigl( m + \frac{r}{c}\dot{m} \Bigr) 
+ \frac{1}{10} \tau^2  
\Bigl( m^{(2)} + \frac{r}{c} m^{(3)} \Bigr) 
+ \frac{1}{280} \tau^4 
\Bigl( m^{(4)} + \frac{r}{c} m^{(5)} \Bigr) 
+ \frac{1}{15120} \tau^6
\Bigl( m^{(6)} + \frac{r}{c} m^{(7)} \Bigr) 
+ \cdots \biggr]_{t - r/c}, 
\label{Gamma_out} 
\end{equation} 
where $m$ and its derivatives are now evaluated at the time
$t-r/c$. It is easy to verify that $\Gamma$ is continuous at
$r=R$. Inserting Eqs.~(\ref{Gamma_in}) and (\ref{Gamma_out}) within
Eq.~(\ref{A_result}) gives us the vector potential of a rotating
charged sphere. 

With the potentials in hand it is a straightforward exercise to obtain
the fields. The electric field can be decomposed into a Coulomb piece
defined by $\bm{E}^{\rm C} = -\bm{\nabla} \Phi$ and an induction piece  
defined by $\bm{E}^{\rm I} = -\partial_t \bm{A}$. These are given by 
\begin{subequations} 
\label{Ein} 
\begin{align} 
\bm{E}^{\rm C}_{\rm in} &= 0, \\ 
\bm{E}^{\rm I}_{\rm in} &= -\frac{\mu_0}{4\pi} \frac{r}{R^3} 
\biggl[ \bigl( \dot{m} + \tau \ddot{m} \bigr) 
+ \frac{1}{10} \frac{r^2}{c^2} \bigl( m^{(3)} + \tau m^{(4)} \bigr) 
+ \frac{1}{280} \frac{r^4}{c^4} \bigl( m^{(5)} + \tau m^{(6)} \bigr) 
\nonumber \\ & \quad \mbox{} 
+ \frac{1}{15120} \frac{r^6}{c^6} \bigl( m^{(7)} + \tau m^{(8)} \bigr) 
+ \cdots \biggr]_{t-R/c} \sin\theta\, \unitphi
\end{align} 
\end{subequations} 
inside the sphere; we recall that $\tau = R/c$. Outside the sphere we
have  
\begin{subequations} 
\label{Eout} 
\begin{align} 
\bm{E}^{\rm C}_{\rm out} &= \frac{1}{4\pi \epsilon_0} \frac{q}{r^2}\, \unitr, \\ 
\bm{E}^{\rm I}_{\rm out} &= -\frac{\mu_0}{4\pi} \frac{1}{r^2} 
\biggl[ \Bigl( \dot{m} + \frac{r}{c} \ddot{m} \Bigr) 
+ \frac{1}{10} \tau^2  
\Bigl( m^{(3)} + \frac{r}{c} m^{(4)} \Bigr) 
+ \frac{1}{280} \tau^4 
\Bigl( m^{(5)} + \frac{r}{c} m^{(6)} \Bigr) 
\nonumber \\ & \quad \mbox{} 
+ \frac{1}{15120} \tau^6
\Bigl( m^{(7)} + \frac{r}{c} m^{(8)} \Bigr) 
+ \cdots \biggr]_{t-r/c} \sin\theta\, \unitphi. 
\end{align} 
\end{subequations} 

The magnetic field $\bm{B} = \bm{\nabla} \times \bm{A} 
= B_r\, \unitr + B_\theta\, \unittheta$ has the
nonvanishing components   
\begin{subequations} 
\label{Bin} 
\begin{align}
B^{\rm in}_r &= \frac{\mu_0}{4\pi} \frac{2}{R^3} 
\biggl[ \bigl( m + \tau \dot{m} \bigr) 
+ \frac{1}{10} \frac{r^2}{c^2} \bigl( m^{(2)} + \tau m^{(3)} \bigr) 
+ \frac{1}{280} \frac{r^4}{c^4} \bigl( m^{(4)} + \tau m^{(5)} \bigr) 
\nonumber \\ & \quad \mbox{} 
+ \frac{1}{15120} \frac{r^6}{c^6} \bigl( m^{(6)} + \tau m^{(7)} \bigr) 
+ \cdots \biggr]_{t-R/c} \cos\theta, \\ 
B^{\rm in}_\theta &= -\frac{\mu_0}{4\pi} \frac{2}{R^3} 
\biggl[ \bigl( m + \tau \dot{m} \bigr) 
+ \frac{1}{5} \frac{r^2}{c^2} \bigl( m^{(2)} + \tau m^{(3)} \bigr) 
+ \frac{3}{280} \frac{r^4}{c^4} \bigl( m^{(4)} + \tau m^{(5)} \bigr) 
\nonumber \\ & \quad \mbox{} 
+ \frac{1}{3780} \frac{r^6}{c^6} \bigl( m^{(6)} + \tau m^{(7)} \bigr) 
+ \cdots \biggr]_{t-R/c} \sin\theta
\end{align} 
\end{subequations} 
inside the sphere, and 
\begin{subequations} 
\label{Bout} 
\begin{align}
B^{\rm out}_r &= \frac{\mu_0}{4\pi} \frac{2}{r^3} 
\biggl[ \Bigl( m + \frac{r}{c} \dot{m} \Bigr) 
+ \frac{1}{10} \tau^2  
\Bigl( m^{(2)} + \frac{r}{c} m^{(3)} \Bigr) 
+ \frac{1}{280} \tau^4 
\Bigl( m^{(4)} + \frac{r}{c} m^{(5)} \Bigr) 
\nonumber \\ & \quad \mbox{} 
+ \frac{1}{15120} \tau^6
\Bigl( m^{(6)} + \frac{r}{c} m^{(7)} \Bigr) 
+ \cdots \biggr]_{t-r/c} \cos\theta, \\ 
B^{\rm out}_\theta &= \frac{\mu_0}{4\pi} \frac{1}{r^3} 
\biggl[ \Bigl( m + \frac{r}{c} \dot{m}  + \frac{r^2}{c^2} \ddot{m} \Bigr) 
+ \frac{1}{10} \tau^2  
\Bigl( m^{(2)} + \frac{r}{c} m^{(3)} + \frac{r^2}{c^2} m^{(4)} \Bigr) 
+ \frac{1}{280} \tau^4 
\Bigl( m^{(4)} + \frac{r}{c} m^{(5)} + \frac{r^2}{c^2} m^{(6)} \Bigr) 
\nonumber \\ & \quad \mbox{} 
+ \frac{1}{15120} \tau^6
\Bigl( m^{(6)} + \frac{r}{c} m^{(7)} + \frac{r^2}{c^2} m^{(8)} \Bigr) 
+ \cdots \biggr]_{t-r/c} \sin\theta 
\end{align} 
\end{subequations} 
outside the sphere. 

\section{Self-torque} 
\label{sec:self-torque}  

The torque exerted by the electromagnetic field on the spinning sphere
--- the self-torque --- is given by Eq.~(\ref{torque_em}), in which we
insert the Lorentz force density of Eq.~(\ref{force_em}). The
calculation requires some care because the fields are discontinuous at
$r=R$. It is easy to see, however, that the Coulomb piece of the
electric field (which is discontinuous) does not contribute to the
torque. The induction piece does contribute, and this contribution is
unambiguous because $\bm{E}^{\rm I}$ happens to be continuous at
$r=R$. The magnetic field is discontinuous, but calculation shows that
the magnetic field makes no contribution to the torque.  

We find after a straightforward calculation that the self-torque's
only nonvanishing component is 
\begin{equation} 
{\scrpt T}^{\rm em}_z = -\frac{\mu_0 q}{6\pi R} 
\biggl[ \bigl( \dot{m} + \tau \ddot{m} \bigr) 
+ \frac{1}{10} \tau^2 \bigl( m^{(3)} + \tau m^{(4)} \bigr) 
+ \frac{1}{280} \tau^4 \bigl( m^{(5)} + \tau m^{(6)} \bigr) 
+ \frac{1}{15120} \tau^6 \bigl( m^{(7)} + \tau m^{(8)} \bigr) 
+ \cdots \biggr]_{t-R/c}. 
\label{selftorque1} 
\end{equation} 
An alternative expression, with $m$ and its derivatives further
expanded about the time $t$, is 
\begin{equation} 
{\scrpt T}^{\rm em}_z = -\frac{\mu_0 q}{6\pi R} 
\biggl[ \dot{m} - \frac{2}{5} \tau^2 m^{(3)} 
+ \frac{1}{3} \tau^3 m^{(4)}
- \frac{6}{35} \tau^4 m^{(5)}  
+ \frac{1}{15} \tau^5 m^{(6)}    
- \frac{4}{189} \tau^6 m^{(7)} 
+ \frac{1}{175} \tau^7 m^{(8)} + \cdots \biggr]_t. 
\label{selftorque2} 
\end{equation}  
The expression of Eq.~(\ref{selftorque1}) is useful for the purpose of
establishing angular momentum balance, as we do in
Sec.~\ref{sec:angmoment}. Equation (\ref{selftorque2}) is a better
starting point for the calculation of the angular motion of the
spinning shell, which we present in Sec.~\ref{sec:motion}. It is also
more convenient to single out the contributions associated with
radiation reaction, as we do in the next paragraphs.    

One would normally expect the self-force or self-torque acting on
a distribution of charge to be associated with the electromagnetic
radiation emitted by the distribution, and one would attach
the words ``radiation reaction'' to the phenomenon. Our result for the
self-torque reveals, however, that this expectation is not entirely
fulfilled: the self-torque is {\it not} the sole result of radiation
reaction. In fact, while radiation reaction does contribute, it is not
the dominant contribution.  

The radiation-reaction terms in Eq.~(\ref{selftorque2}) can be
identified as those that come with an odd power of $c$ (recall that
$\tau = R/c$); the terms that come with even powers of $c$ have
nothing to do with radiation reaction. To understand this, recall that
the fields of Sec.~\ref{sec:fields} were calculated on the basis of
the retarded Green's function. These fields describe an outgoing
electromagnetic wave that carries angular momentum outward, with the
radiated loss of angular momentum balanced by a contribution to the 
self-torque. Suppose, however, that we had adopted the advanced
solution instead of the retarded solution. In this case we would
have had an incoming wave that carries angular momentum inward, and
the radiated {\it gain} of angular momentum would have been balanced
by an equal and opposite contribution to the self-torque. The advanced
solution can be obtained from the retarded solution by the formal
switch $c \to -c$, and the contributions to the self-torque that are
sensitive to the choice of solution are therefore those that are odd
in $c$; these are the radiation-reaction terms. The terms that are
even in $c$ are the same regardless of the choice of solution, and
these contributions to the self-torque --- though they still have to 
do with the electromagnetic field created by the spinning sphere ---
have nothing to do with radiation reaction.  

Another way of stating all this is to say that the complete
self-torque is a sum of time-antisymmetric terms (odd in $c$) that
make up the radiation-reaction piece, and time-symmetric terms (even
in $c$) that have nothing to do with radiation reaction.  The notions
of self-torque and radiation reaction must therefore be distinguished.   

Inspection of Eq.~(\ref{selftorque2}) reveals that the leading terms
involving $\dot{m}$ and $m^{(3)}$ come with even powers of $c$ and are
therefore time-symmetric; these contributions to the self-torque have
nothing to do with radiation reaction. The leading radiation-reaction
term occurs at order $\tau^3$ and is proportional to $m^{(4)}$.  

\section{Radiated angular momentum}  
\label{sec:radiated} 

The radiated flux of angular momentum is calculated on the basis of
Eq.~(\ref{flux}). We choose $S$ to be a sphere of radius $r_0$, and we
take $r_0$ to be very large compared with $\lambda_c$, a
characteristic wavelength of the electromagnetic radiation. With $t_c$  
denoting the characteristic time scale of variation of the magnetic
moment (this quantity was introduced in Sec.~\ref{sec:fields}), we
have that $\lambda_c = c t_c$, and we therefore assume that 
$r_0 \gg c t_c \gg R$. 

After inserting Eq.~(\ref{maxwell_stress}) and writing 
$dS_a = \hat{r}_a r_0^2\, d\Omega$ for the surface element (where
$d\Omega := \sin\theta\, d\theta d\phi$), we find that the
angular-momentum flux can be expressed as 
\begin{equation} 
{\scrpt N}^{\rm em}_a = -\epsilon_0  
\int r_0^3 (\unitr \cdot \bm{E}) (\unitr \times \bm{E})_a\, d\Omega 
- \frac{1}{\mu_0}  
\int r_0^3 (\unitr \cdot \bm{B}) (\unitr \times \bm{B})_a\, d\Omega. 
\end{equation} 
We make the substitutions from Eqs.~(\ref{Eout}) and (\ref{Bout}),
neglect terms that are suppressed by powers of $\lambda_c/r_0$, perform
the angular integration, and obtain that the only nonvanishing
component of the radiated flux is 
\begin{equation} 
{\scrpt N}^{\rm em}_z = \frac{\mu_0 q}{6\pi c}\biggl[ \ddot{m} 
+ \frac{1}{10} \tau^2 m^{(4)} 
+ \frac{1}{280} \tau^4 m^{(6)} 
+ \frac{1}{15120} \tau^6 m^{(8)} + \cdots \biggr]_{t-r_0/c}, 
\label{radiatedflux} 
\end{equation} 
where $m$ and its derivatives are evaluated at the time $t - r_0/c$; 
we recall that $\tau = R/c$. The expression is proportional to $q$ 
and linear in the magnetic moment; this indicates that the radiated
angular momentum results from an interplay between the Coulomb and
induction pieces of the electric field.  

The radiated flux of Eq.~(\ref{radiatedflux}) is associated with the
retarded solution to Maxwell's equation. As was discussed in
Sec.~\ref{sec:self-torque}, the advanced solution can be obtained with
the formal switch $c \to -c$, and the advanced version of the radiated
flux would come with an overall minus sign, and would involve $m$ and
its derivatives evaluated at the time $t + r_0/c$.   

\section{Field angular momentum} 
\label{sec:angmoment} 

The self-torque of Eq.~(\ref{selftorque1}) and the radiated angular
momentum of Eq.~(\ref{radiatedflux}) come at the expense of the
field's own angular momentum, as implied by
Eq.~(\ref{conservation_em}). The angular momentum in the
electromagnetic field can be calculated from Eq.~(\ref{em_angmom}),
which we write in the form 
\begin{equation} 
J^{\rm em}_a = \epsilon_0 \int_V \bigl[ (\bm{x} \cdot \bm{B}) E_a
- (\bm{x} \cdot \bm{E}) B_a \bigr]\, d^3x, 
\end{equation} 
or 
\begin{equation} 
J^{\rm em}_a = \int _0^{r_0} \gothj_a\, dr 
\end{equation} 
with 
\begin{equation} 
\gothj_a := \epsilon_0 r^3 \int \bigl[ (\unitr \cdot \bm{B}) E_a
- (\unitr \cdot \bm{E}) B_a \bigr]\, d\Omega; 
\end{equation} 
we recall that $r_0$ is the radius of the sphere $S$ bounding $V$. 
Substitution from Eqs.~(\ref{Ein}) and (\ref{Bin}) and evaluation of
the angular integral reveals that $\gothj_a = 0$ when $r < R$. For 
$r > R$ we get from Eqs.~(\ref{Eout}) and (\ref{Bout}) that the only 
nonvanishing component of $\gothj_a$ is  
\begin{align} 
\gothj_z &= \frac{\mu_0 q}{6\pi} \biggl[ 
\biggl( \frac{m}{r^2} + \frac{\dot{m}}{cr} + \frac{\ddot{m}}{c^2} \biggr)  
+ \frac{1}{10} \tau^2  
\biggl( \frac{m^{(2)}}{r^2} + \frac{m^{(3)}}{cr} + \frac{m^{(4)}}{c^2} \biggr)   
+ \frac{1}{280} \tau^4 
\biggl( \frac{m^{(4)}}{r^2} + \frac{m^{(5)}}{cr} + \frac{m^{(6)}}{c^2} \biggr)   
\nonumber \\ & \quad \mbox{} 
+ \frac{1}{15120} \tau^6
\biggl( \frac{m^{(6)}}{r^2} + \frac{m^{(7)}}{cr} + \frac{m^{(8)}}{c^2} \biggr)   
+ \cdots \biggr]_{t-r/c}. 
\end{align} 
This must be integrated with respect to $r$, and it is a fortunate
circumstance that each term in $\gothj_z$ is a derivative with respect
to $r$. For example, 
\begin{equation} 
\frac{m}{r^2} + \frac{\dot{m}}{cr} + \frac{\ddot{m}}{c^2}  
= -\frac{\partial}{\partial r} \biggl( \frac{m}{r} + \frac{\dot{m}}{c}
\biggr), 
\end{equation} 
where the partial derivative holds $t$ fixed. All remaining terms can
be written in a similar way. The integration is performed from $r=R$
to $r=r_0$, and after neglecting terms that are suppressed by powers
of $\lambda_c/r_0$, we arrive at   
\begin{align} 
J^{\rm em}_z &= \frac{\mu_0 q}{6\pi R} 
\biggl[ \bigl( m + \tau \dot{m} \bigr) 
+ \frac{1}{10} \tau^2 \bigl( m^{(2)} + \tau m^{(3)} \bigr) 
+ \frac{1}{280} \tau^4 \bigl( m^{(4)} + \tau m^{(5)} \bigr) 
+ \frac{1}{15120} \tau^6 \bigl( m^{(6)} + \tau m^{(7)} \bigr) 
+ \cdots \biggr]_{t-R/c} 
\nonumber \\ & \quad \mbox{} 
- \frac{\mu_0 q}{6\pi c} \biggl[ \dot{m} 
+ \frac{1}{10} \tau^2 m^{(3)} 
+ \frac{1}{280} \tau^4 m^{(5)} 
+ \frac{1}{15120} \tau^6 m^{(7)} + \cdots \biggr]_{t-r_0/c}. 
\label{emJresult}
\end{align} 
As indicated, the first set of terms is evaluated at the time
$t-R/c$, while the second set is evaluated at $t-r_0/c$. The
first set can be thought of as the contribution to the
angular momentum coming from the near-zone fields in the vicinity of
the spinning shell. The second set can be thought of as the
contribution from the wave-zone fields.  

Differentiation of Eq.~(\ref{emJresult}) with respect to $t$ and
comparison with Eqs.~(\ref{selftorque1}) and (\ref{radiatedflux})
reveals that the field angular momentum satisfies 
$d J^{\rm em}_a/dt = -{\scrpt N}^{\rm em}_a - {\scrpt T}^{\rm em}_a$.  
This is Eq.~(\ref{conservation_em}), the statement of angular momentum
balance that was first established in Sec.~\ref{sec:conservation}. It
is deeply satisfying that the statement can be verified explicitly.   

\section{Shell motion} 
\label{sec:motion} 

The self-torque of Eq.~(\ref{selftorque2}) affects the angular motion
of the shell. The motion is governed by Eq.~(\ref{conservation_mech}),
$dJ^{\rm mech}_a/dt = {\scrpt T}_a^{\rm ext} 
+ {\scrpt T}_a^{\rm em}$, where $J^{\rm mech}_a$ is the shell's
mechanical angular momentum, ${\scrpt T}_a^{\rm ext}$ the torque
supplied by an external agent, and ${\scrpt T}_a^{\rm em}$ the
self-torque. In this section we explore the consequences of this
equation. 

\subsection{Equation of angular motion}   

We take the shell to have a uniform mass density, and its mechanical
angular momentum is related to the magnetic moment by 
\begin{equation} 
\bm{J}^{\rm mech} = \frac{2M}{q} \bm{m}, 
\end{equation} 
where $M$ is the shell's total mass. The shell's equation of motion is
therefore 
\begin{equation} 
\frac{2M}{q} \dot{m} = {\scrpt T}^{\rm ext} 
- \frac{\mu_0 q}{6\pi R} 
\biggl( \dot{m} - \frac{2}{5} \tau^2 m^{(3)} 
+ \frac{1}{3} \tau^3 m^{(4)}
- \frac{6}{35} \tau^4 m^{(5)}  
+ \frac{1}{15} \tau^5 m^{(6)}    
- \frac{4}{189} \tau^6 m^{(7)} 
+ \frac{1}{175} \tau^7 m^{(8)} + \cdots \biggr),
\end{equation} 
where ${\scrpt T}^{\rm ext}$ is the $z$-component of the external
torque (we assume that the remaining components vanish). We transfer
the $\dot{m}$ term on the right-hand side of the equation to the other
side, and write the combination as $(2M_{\rm eff}/q) \dot{m}$, where 
$M_{\rm eff} := M + \delta M$ is an effective mass, with an
electrostatic correction given by 
\begin{equation} 
\delta M c^2 = \frac{q^2}{12\pi\epsilon_0 R}. 
\label{notation1} 
\end{equation} 
This contribution to the effective mass can be thought of as 
additional inertia provided by the electromagnetic field. Curiously,
$\delta M c^2$ is a factor $2/3$ smaller than the electrostatic energy
of a hollow sphere, given by $U = q^2/(8\pi\epsilon_0 R)$.  

Introducing the notation 
\begin{equation} 
{\sf T} := \frac{q}{2M_{\rm eff}} {\scrpt T}^{\rm ext}, \qquad 
\varepsilon := \frac{1}{M_{\rm eff} c^2} 
\frac{q^2}{30\pi \epsilon_0 R} = \frac{2}{5} \frac{\delta M}{M_{\rm eff}}, 
\label{notation2} 
\end{equation} 
we put the shell's equation of motion in the form   
\begin{equation} 
\dot{m} = {\sf T} + \varepsilon \biggl( \tau^2 m^{(3)} 
- \frac{5}{6} \tau^3 m^{(4)}
+ \frac{3}{7} \tau^4 m^{(5)}  
- \frac{1}{6} \tau^5 m^{(6)}    
+ \frac{10}{189} \tau^6 m^{(7)} 
- \frac{1}{70} \tau^7 m^{(8)} + \cdots \biggr), 
\label{EoM} 
\end{equation} 
where, we recall, $\tau := R/c$. We note that up to a factor of order
unity, $\varepsilon$ is the ratio of the electrostatic contribution to
the mass to the total effective mass. This ratio is small for a
macroscopic body. To take the measure of this, we consider a
sphere of mass $M = 10\ \mbox{g}$ and radius $R = 1\ \mbox{cm}$, on
which we deposit a charge $q = 0.1\ \mbox{C}$; for this object 
$\varepsilon \simeq 1 \times 10^{-6}$. The assumption
placed on the variation of $m$, $\tau/t_c \ll 1$, ensures that 
$\tau^2 m^{(3)} \ll \dot{m}$, 
$\tau^3 m^{(4)} \ll \tau^2 m^{(3)}$, and so on. The upshot is that the 
external term ${\sf T}$ is strongly dominant in Eq.~(\ref{EoM}), and 
the self-torque terms provide small corrections. We recall that in
Eq.~(\ref{EoM}), the terms with odd powers of $\tau$ are
radiation-reaction terms; the terms with even powers of $\tau$
(including the leading-order term proportional to $m^{(3)}$) have
nothing to do with radiation reaction.  

Equation (\ref{EoM}) contains high-order derivatives, and it would
include terms of even higher order if the expansion in powers of 
$\tau/t_c$ were pursued beyond what is displayed there. Such
equations are typically pathological and produce runaway solutions
that strongly violate the underlying assumption that 
$\tau/t_c \ll 1$. A coping strategy presents itself on the grounds
that $\varepsilon$ is small, and that each successive term contributes
a smaller correction. The strategy is to perform a reduction of order,
by substituting $\dot{m} = {\sf T} + O(\varepsilon)$ on the right-hand 
side of the equation. This yields   
\begin{equation} 
\dot{m} = {\sf T} + \varepsilon \biggl( \tau^2 {\sf T}^{(2)} 
- \frac{5}{6} \tau^3 {\sf T}^{(3)}
+ \frac{3}{7} \tau^4 {\sf T}^{(4)}  
- \frac{1}{6} \tau^5 {\sf T}^{(5)}    
+ \frac{10}{189} \tau^6 {\sf T}^{(6)} 
- \frac{1}{70} \tau^7 {\sf T}^{(7)} + \cdots \biggr), 
\label{EoM_reduced} 
\end{equation} 
where we neglect all terms of order $\varepsilon^2$ and
higher.\footnote{The reduction of order works even when $\varepsilon$
  is taken to be of order unity. In this case the procedure relies
  entirely on the assumption that $\tau/t_c \ll 1$, and it produces a
  very similar form for the reduced equation of motion, with
  coefficients that acquire $\varepsilon$-dependent corrections.} 

Equation (\ref{EoM_reduced}) is free of pathologies, and its solution
is immediate: 
\begin{equation} 
m(t) = \int^t {\sf T}(t')\, dt' 
+ \varepsilon \biggl( \tau^2 {\sf T}^{(1)} 
- \frac{5}{6} \tau^3 {\sf T}^{(2)}
+ \frac{3}{7} \tau^4 {\sf T}^{(3)}  
- \frac{1}{6} \tau^5 {\sf T}^{(4)}    
+ \frac{10}{189} \tau^6 {\sf T}^{(5)} 
- \frac{1}{70} \tau^7 {\sf T}^{(6)} + \cdots \biggr)
+ \mbox{constant}, 
\label{EoM_solution} 
\end{equation} 
where the constant is determined by the initial condition. With 
${\sf T}$ assumed to vary on a time scale $t_c$ that is long 
compared with $\tau = R/c$, Eq.~(\ref{EoM_solution}) is guaranteed to 
describe a $m(t)$ that also varies on a long time scale.    

\subsection{Examples} 

For the purpose of illustration we examine a number of concrete
examples. 

To begin we consider the simplest case, that of a vanishing external
torque. Setting ${\sf T} = {\sf T}_1 := 0$ in Eq.~(\ref{EoM_solution})
returns $m_1(t) = \mbox{constant}$, and we see that as expected, the
sphere rotates with a constant angular velocity when the external
torque is zero. Equation (\ref{selftorque2}) implies that the self-torque
vanishes, Eq.~(\ref{radiatedflux}) shows that there is no radiated
flux of angular momentum, and Eq.~(\ref{emJresult}) reveals that the
field's angular momentum is $J^{\rm em} = \mu_0 q m_1 / (6\pi R)$, a
constant. In this rather trivial case, angular momentum is strictly
conserved.   

Next on our list of examples is that of a constant external torque, 
${\sf T} = {\sf T}_2 := \mbox{constant}$. In this case
Eq.~(\ref{EoM_solution}) returns $m_2(t) = m_2(0) + {\sf T}_2\, t$, and we
find that the angular velocity increases linearly with time. Equation
(\ref{selftorque2}) implies that the self-torque is given by 
\begin{equation} 
{\scrpt T}^{\rm em} = -\frac{\mu_0 q}{6\pi R} {\sf T}_2
= -\frac{\delta M}{M_{\rm eff}} {\scrpt T}^{\rm self}
= -\frac{5}{2} \varepsilon\, {\scrpt T}^{\rm self}, 
\end{equation} 
where we used Eqs.~(\ref{notation1}) and (\ref{notation2}). This
self-torque was already moved to the left-hand side of the equation of
motion in Eq.~(\ref{EoM}), and shown to give rise to the shift $M \to
M + \delta M = M_{\rm eff}$ in the sphere's mass. The previous 
equation reveals another interpretation for the self-torque
(valid in this specific case only): it produces a slight offset
in the external torque. Equation (\ref{radiatedflux}) shows that there
is no radiated flux of angular momentum.   

Next we consider an external torque given by 
\begin{equation} 
{\sf T} = {\sf T}_3 := \gothm k^3 t^2,
\end{equation} 
where $\gothm$ is a constant with unit of magnetic moment, and $k$ a 
constant with unit of inverse time. The solution to
Eq.~(\ref{EoM_solution}) is 
\begin{equation} 
m_3(t) = m_3(0) + \frac{1}{3} \gothm (kt)^3 
+ 2\varepsilon \gothm \zeta^2 (k t), 
\end{equation} 
where $\zeta := k\tau \ll 1$. In this case the magnetic moment is a
cubic function of $t$. The leading term, growing as $t^3$,
is contributed by the external torque, and the subleading term,
proportional to $\varepsilon$ and growing as $t$, is contributed by
the self-torque. It is interesting to note that the self-torque term
is proportional to $\zeta^2 \propto \tau^2 \propto c^{-2}$ and
therefore even in $c$. Recalling the discussion of
Sec.~\ref{sec:self-torque}, we conclude that the self-torque
correction has nothing to do with radiation
reaction. This is in spite of the fact that a magnetic moment that
grows approximately as $t^3$ does radiate angular momentum, as shown
by Eq.~(\ref{radiatedflux}). The apparent paradox is resolved by
noting that the radiated losses come at the expense of the field's
angular momentum, and need not be balanced by a radiation-reaction
contribution to the self-torque.     

As a slight variation of the preceding example, we take 
\begin{equation} 
{\sf T} = {\sf T}_4 := \gothm k^4 t^3, 
\end{equation} 
where $\gothm$ and $k$ are again constants. The corresponding solution
to Eq.~(\ref{EoM_solution}) is  
\begin{equation} 
m_4(t) = m_4(0) + \frac{1}{4} \gothm (k t)^4 + \varepsilon \gothm 
\Bigl[ 3 \zeta^2 (kt)^2 - 5 \zeta^3 (kt) \Bigr], 
\end{equation} 
where $\zeta := k\tau \ll 1$. The leading term that grows as $t^4$ is
contributed by the external torque, and the remaining terms are
contributed by the self-torque. In this case we see that the
self-torque correction contains a term proportional to $\zeta^2$,
which is again even in $c$; this term has
nothing to do with radiation reaction. But there is also a term
proportional to $\zeta^3$, which is odd in $c$; this is a
radiation-reaction contribution.  

Next we consider an oscillating external torque given by 
\begin{equation} 
{\sf T} = {\sf T}_5 := \gothm \omega \cos\omega t, 
\end{equation} 
where $\gothm$ is a constant with unit of magnetic moment, and
$\omega$ is a constant angular frequency. The corresponding solution 
to the equation of motion is  
\begin{align} 
m_5(t) &= m_5(0) + \gothm \biggl[ 1 - \varepsilon \biggl( 
\zeta^2 - \frac{3}{7} \zeta^4 + \frac{10}{189} \zeta^6 + \cdots \biggr)
\biggr] \sin\omega t 
\nonumber \\ & \quad \mbox{} 
+ \varepsilon \gothm \biggl( \frac{5}{6} \zeta^3 
- \frac{1}{6} \zeta^5 + \frac{1}{70} \zeta^7 + \cdots \biggr) 
\bigl( \cos\omega t - 1 \bigr), 
\end{align} 
where $\zeta := \omega \tau \ll 1$. In the first set of terms,
proportional to $\sin\omega t$, the self-torque produces an amplitude 
correction that is even in $\zeta$; this correction is not to be
associated with radiation reaction. The second set of terms is also 
contributed by the self-torque, and it represents a correction to
the phase of the magnetic moment; these terms are odd in $\zeta$, and
they are directly associated with radiation reaction. 

For a final example we consider an external torque that gradually
switches on and then gradually switches off. We describe it in terms
of a gaussian function, 
\begin{equation} 
{\sf T} = {\sf T}_6 := \gothm k e^{-k^2 t^2}, 
\end{equation} 
so that the external torque is active during a time interval
proportional to $k^{-1}$ around $t=0$. The corresponding solution is 
\begin{align} 
m_6(t) &= m_6(-\infty) 
+ \frac{1}{2} \sqrt{\pi} \gothm \bigl[ \mbox{erf}(kt) + 1 \bigr] 
- \varepsilon \gothm \biggl\{ 2 \zeta^2 (kt) 
+ \frac{5}{3} \zeta^3 \bigl[ 2(kt)^2 - 1 \bigr]  
+ \frac{12}{7} \zeta^4 \bigl[ 2(kt)^3 - 3(kt) \bigr] 
\nonumber \\ & \quad \mbox{}
+ \frac{2}{3} \zeta^5 \bigl[ 4(kt)^4 - 12(kt)^2 + 3 \bigr] 
+ \frac{80}{189} \zeta^6 \bigl[ 4(kt)^5 - 20(kt)^3 + 15(kt) \bigr] 
\nonumber \\ & \quad \mbox{}
+ \frac{4}{35} \zeta^7 \bigl[ 8(kt)^6 - 60(kt)^4 + 90(kt)^2 
-15 \bigr] + \cdots \biggr\} e^{-k^2 t^2}, 
\label{m4} 
\end{align} 
where $\mbox{erf}(u) := 2 \pi^{-1/2} \int_0^u e^{-v^2}\, dv$ is the
error function. The leading term in Eq.~(\ref{m4}) describes a gradual
increase of $m_6$ from its initial value $m_6(-\infty)$ to its final value
$m_6(+\infty) = m_6(-\infty) + \sqrt{\pi} \gothm$. The corrections
contributed by the self-torque are slight modulations superposed to
the leading behavior described by the error function.  

\begin{acknowledgments} 
Conversations with Bernie Nickel were greatly appreciated. 
This work was supported by the Natural Sciences and Engineering
Research Council of Canada and by Perimeter Institute for Theoretical
Physics. Research at Perimeter Institute is supported by the
Government of Canada through the Department of Innovation, Science and
Economic Development Canada and by the Province of Ontario through the
Ministry of Research, Innovation and Science.   
\end{acknowledgments} 

\appendix 

\section{Spherical-harmonic decomposition of the retarded Green's
  function} 

The retarded Green's function for the wave equation is given by 
\begin{equation}
G(t,\bm{x};t',\bm{x'}) 
= \frac{\delta(t-t'-|\bm{x}-\bm{x'}|/c)}{|\bm{x}-\bm{x'}|}. 
\end{equation}
Alternatively, it can be expressed as 
\begin{equation} 
G(t,\bm{x};t',\bm{x'}) = 2c\Theta(t-t')\, \delta\bigl[ c^2(t-t')^2  
- |\bm{x}-\bm{x'}|^2 \bigr], 
\label{G} 
\end{equation} 
with $\Theta(u)$ denoting the step function. 

We wish to express the Green's function as a spherical-harmonic
decomposition. We write  
\begin{equation}
G(t,\bm{x};t',\bm{x'}) = \sum_{\ell \m} g_\ell(t,r;t',r') 
Y^*_{\ell \m}(\theta',\phi') Y_{\ell \m}(\theta,\phi), 
\label{G_decomposed} 
\end{equation}
where $g_\ell(t,r;t',r')$ is a reduced Green's function for
each multipole order $\ell$; the label $\m$ does not appear because the  
reduced wave equation satisfied by $g_\ell$ is independent of $\m$.  
The strategy to obtain $g_\ell$ is to subject Eq.~(\ref{G}) to a
projection to each one of its multipole components. 

We introduce the notation $\Delta := c(t-t')$, $s :=
|\bm{x}-\bm{x'}|$, and express Eq.~(\ref{G}) in the condensed form   
$G = 2c \Theta(\Delta) \delta(\Delta^2 - s^2)$. We substitute this on
the left-hand side of Eq.~(\ref{G_decomposed}), multiply each side by
$Y_{\ell' \m'}(\theta',\phi')$, and integrate over 
$d\Omega' = \sin\theta'\, d\theta' d\phi'$. The result is 
\begin{equation} 
2c \Theta(\Delta) \int \delta(\Delta^2 - s^2) 
Y_{\ell \m}(\theta',\phi')\, d\Omega' 
= g_\ell Y_{\ell \m}(\theta,\phi). 
\end{equation} 
We next set $\m=0$ and use the fact that $Y_{\ell 0}(\theta,\phi) 
\propto P_\ell(\cos\theta)$. The previous equation reduces to 
\begin{equation} 
2c \Theta(\Delta) \int \delta(\Delta^2 - s^2) 
P_\ell(\cos\theta')\, d\cos\theta'\, d\phi'  
= g_\ell P_\ell (\cos\theta).
\end{equation} 
Finally, we set $\cos\theta = 1$ and use the fact that 
$P_\ell(1) = 1$. This gives 
\begin{equation} 
g_\ell = 2 c \Theta(\Delta) \int \delta(\Delta^2 - s^2)
\Bigr|_{\cos\theta = 1} P_\ell(\cos\theta')\, d\cos\theta'\, d\phi'. 
\end{equation} 
Because $s^2$ evaluated at $\cos\theta = 1$ is
$r^2 - 2 r r' \cos\theta' + r^{\prime 2}$, we have that  
\begin{equation} 
g_\ell = 4\pi c \Theta(\Delta) \int
\delta(\Delta^2 - r^2 + 2 r r' \cos\theta' - r^{\prime 2}) 
P_\ell(\cos\theta')\, d\cos\theta', 
\end{equation} 
or 
\begin{equation} 
g_\ell = \frac{2\pi c\Theta(\Delta)}{rr'} \int 
\delta(\cos\theta' - \xi) P_\ell(\cos\theta')\, d\cos\theta', 
\end{equation} 
where $\xi := (r^2 + r^{\prime 2} - \Delta^2)/(2rr')$. 

The integral is nonzero whenever $\xi$ lies in the interval between
$-1$ and $+1$; when this condition is satisfied it evaluates to 
$g_\ell = 2\pi c\Theta(\Delta) P_\ell(\xi)/(rr')$.   
The condition $-1 < \xi$ implies $-2rr' < r^2 + r^{\prime 2}
- \Delta^2$, so that $\Delta < r + r'$. The condition $\xi < 1$
implies $2rr' > r^2 + r^{\prime 2} - \Delta^2$, so that 
$\Delta > |r - r'|$. This last condition supersedes the requirement 
that $\Delta > 0$, which comes from the step function appearing in
$G$. Altogether, we find that the reduced Green's function is
given by  
\begin{equation} 
g_\ell(t,r;t',r') = \frac{2\pi c}{rr'} 
\Theta(\Delta - |r-r'|) \Theta(r+r' - \Delta) 
P_\ell(\xi). 
\end{equation}
The temporal support of the reduced Green's function is the interval  
$|r-r'| < \Delta < r + r'$.  

\bibliography{../bib/master}

\begin{thebibliography}{10}
\expandafter\ifx\csname url\endcsname\relax
  \def\url#1{{\tt #1}}\fi
\expandafter\ifx\csname urlprefix\endcsname\relax\def\urlprefix{URL }\fi

\bibitem{griffiths:13}
D.~J. Griffiths, {\em Introduction to electrodynamics. Fourth edition\/}
  (Pearson, Boston, 2013).

\bibitem{romer:66}
R.~H. Romer, {\em Angular momentum of static electromagnetic fields\/}, Am. J.
  Phys. {\bf 34}, 772--778 (1966).

\bibitem{boos:84}
F.~L. Boos, Jr, {\em More on the Feynman's disk paradox\/}, Am. J. Phys. {\bf
  52}, 756--757 (1984).

\bibitem{feynman-leighton-sands:11}
R.~P. Feynman, R.~B. Leighton, and M.~Sands, {\em The Feynman Lectures on
  Physics, The New Millennium Edition\/} (Basic Books, USA, 2011).

\bibitem{lombardi:83}
G.~G. Lombardi, {\em Feynman's disk paradox\/}, Am. J. Phys. {\bf 51}, 213--214
  (1983).

\bibitem{jackson:99}
J.~D. Jackson, {\em Classical Electrodynamics, Third Edition\/} (Wiley, New
  York, 1999).

\bibitem{boyer:72}
T.~H. Boyer, {\em Mass renormalization and radiation damping for a charged
  particle in uniform circular motion\/}, Am. J. Phys. {\bf 40}, 1843--1846
  (1972).

\bibitem{rohrlich:97}
F.~Rohrlich, {\em The dynamics of a charged sphere and the electron\/}, Am. J.
  Phys. {\bf 65}, 1051--1056 (1997).

\bibitem{griffiths-proctor-schroeter:10}
D.~J. Griffiths, T.~C. Proctor, and D.~F. Schroeter, {\em Abraham--Lorentz
  versus Landau--Lifshitz\/}, Am. J. Phys. {\bf 78}, 391--402 (2010).

\bibitem{landau-lifshitz:b2}
L.~D. Landau and E.~M. Lifshitz, {\em The Classical Theory of Fields, Fourth
  Edition\/} (Butterworth-Heinemann, Oxford, England, 2000).

\bibitem{griffiths-owen:83}
D.~J. Griffiths and R.~E. Owen, {\em Mass renormalization in classical
  electrodynamics\/}, Am. J. Phys. {\bf 51}, 1120--1126 (1983).

\bibitem{griffiths-szeto:78}
D.~J. Griffiths and E.~W. Szeto, {\em Dumbbell model for the classical
  radiation reaction\/}, Am. J. Phys. {\bf 46}, 244--248 (1978).

\bibitem{ori-rosenthal:03}
A.~Ori and E.~Rosenthal, {\em Universal self force from an extended-object
  approach\/}, Phys. Rev. D {\bf 68}, 041701 (2003), arXiv:gr-qc/0205003.

\bibitem{ori-rosenthal:04}
A.~Ori and E.~Rosenthal, {\em Calculation of the self force using the
  extended-object approach\/}, J. Math. Phys. {\bf 45}, 2347--2364 (2004),
  arXiv:gr-qc/0309102.

\bibitem{gralla-harte-wald:09}
S.~E. Gralla, A.~I. Harte, and R.~M. Wald, {\em Rigorous derivation of
  electromagnetic self-force\/}, Phys. Rev. D {\bf 80}, 024031 (2009),
  arXiv:0905.2391.

\bibitem{decastro:91}
A.~S. de~Castro, {\em Electromagnetic angular momentum for a rotating charged
  shell\/}, Am. J. Phys. {\bf 59}, 180--181 (1991).

\bibitem{stump-pollack:97}
D.~R. Stump and G.~L. Pollack, {\em Magnetic dipole oscillations and radiation
  damping\/}, Am. J. Phys. {\bf 65}, 81--87 (1997).

\bibitem{ashtekar-bonga:17}
A.~Ashtekar and B.~Bonga, {\em {On the ambiguity in the notion of transverse
  traceless modes of gravitational waves}\/}, Gen. Rel. Grav. {\bf 49}, 122
  (2017), arXiv:1707.09914.

\bibitem{cook:84}
R.~J. Cook, {\em Radiation reaction revisited\/}, Am. J. Phys. {\bf 52},
  894--895 (1984).

\end{thebibliography}
\end{document}